\documentclass{article}
\usepackage{amsthm,amssymb,amsfonts,url,a4wide}
\usepackage{tikz}
\title{\bf Infinite Secret Sharing -- Examples\thanks{This research has been
partially supported by the ``Lend\"ulet'' Program of the Hungarian Academy
of Sciences.}}
\author{Alexander Dibert$^1$\and L\'aszl\'o Csirmaz$^{1,2}$}
\date{\small $^1$ Central European University, Budapest
\\
     $^2$ Renyi Institute, Budapest}

\newtheorem{theorem}{Theorem}[section]
\newtheorem{definition}[theorem]{Definition}

\newtheorem{scheme}{Scheme}[section]

\newcommand{\restr}{\raise.3ex\hbox{$\upharpoonright$}}
\newcommand\tsum{\mathop{\textstyle\sum}}
\def\eps{\varepsilon}
\def\phi{\varphi}
\def\defeq{\buildrel{\mbox{\footnotesize def}}\over=}

\let\oldtriangle=\triangle
\def\triangle{\mathop{\raise.2ex\hbox{$\scriptstyle\oldtriangle$}}}
\def\Prob{\mathop{\hbox{\rm Prob}}\nolimits}
\makeatletter
\def\pmod#1{\allowbreak\mkern5mu({\operator@font mod}\,\,#1)}
\makeatother

\begin{document}
\maketitle

\begin{abstract}
The motivation for extending secret sharing schemes to cases when either the
set of players is infinite or the domain from which the secret and/or the
shares are drawn is infinite or both, is similar to the case when switching
to abstract probability spaces from classical combinatorial probability. It
might shed new light on old problems, could connect seemingly unrelated
problems, and unify diverse phenomena.

Definitions equivalent in the finitary case could be very much different
when switching to infinity, signifying their difference. The standard
requirement that qualified subsets should be able to determine the secret
has different interpretations in spite of the fact that, by assumption, all
participants have infinite computing power. The requirement that unqualified
subsets should have no, or limited information on the secret suggests that
we also need some probability distribution. In the infinite case events with
zero probability are not necessarily impossible, and we should decide
whether bad events with zero probability are allowed or not.

In this paper, rather than giving precise definitions, we enlist an abundance
of hopefully interesting infinite secret sharing schemes. These
schemes touch quite diverse areas of mathematics such as projective
geometry, stochastic processes and Hilbert spaces. Nevertheless our main 
tools are from probability theory. The examples discussed here serve as
foundation and illustration to the more theory oriented companion paper
\cite{infinite-secret-theory}.


\noindent
{\bf AMS classification numbers:} 60A99, 60B05, 60G15, 62F10, 94A62, 46C99,
54D10

\end{abstract}

\section{Introduction}
The topic of this paper is to provide several examples of secret
sharing schemes where the domain of secret, that of shares, or
the set of players 
is infinite. This type of approach studying infinite objects
instead of finitary ones is not novel even in the realm of
cryptography, see, e.g., \cite{blakley-swanson, ChorKush,
Makar:1980, patarin:infinite, Vaudenay:09}. As usual, switching to
infinite means some kind of abstraction: we disregard particular and quite
frequently annoying properties of finite structures, and focus on their
general properties only. An excellent example is Probability Theory where
the law of large numbers automatically leads to continuous distributions.

The concept of secret sharing either on infinite domain or with infinitely
many players is quite straightforward. Nevertheless, the difficulty lies in
the details. Natural properties fail in the infinite case, and even exact
definitions are sometimes problematic. In this paper we describe several 
natural schemes which are intuitively correct. 
We also give counterexamples
showing that certain ``natural'' definitions might not achieve the 
desired effects.

We mainly consider {\it probabilistic schemes}, where
correctness and completeness relies on some probability measure. Even
defining probability measures on arbitrary (product) space is not without
problems, see \cite{delarue, haezendonck, rokhlin} for a general description
of the problems and the definition for the ``standard probability space.''
While we do not rely on those works, a good working
knowledge of measure theory and probability spaces will definitely help
understanding the basic issues.

Schemes are grouped more or less arbitrarily according to several
contradictory properties: the method used in the scheme, their similarity to
each other, and the access structure they realize. Section
\ref{sec:definitions} enlists the definitions of {\em access structure},
(probabilistic) {\em secret sharing scheme}, and when a scheme {\em
realizes} a structure. Wording of definitions are sometimes vague, as we did
not want to {\em a priori} exclude schemes which otherwise would intuitively
fit into this collection. Exact definitions, as extracted from these
examples, can be found in the companion paper \cite{infinite-secret-theory}.
Section \ref{sec:uniform}--\ref{sec:statistics} contain schemes which
illustrate the diversity of tools they use. As it is proved in
\cite{infinite-secret-theory}, schemes in Sections \ref{sec:stochastic},
\ref{sec:statistics} and \ref{sec:hilbert} are the best possible: no scheme
with better security guarantee can realize the same access structure. Section
\ref{sec:hilbert} introduces the notion of {\em Hilbert space programs}
which is a generalization of span programs \cite{spanprogram} for infinitely
many participants. Section \ref{sec:esoteric} concludes the paper with a
scheme which is not a (probabilistic) scheme at all.

\section{Definitions}\label{sec:definitions}

In this section we define the basic notions of secret sharing, namely what
access structure is, and give an informal definition for 
secret sharing schemes and when it realizes an access structure. 
Precise definitions are postponed to the companion paper
\cite{infinite-secret-theory}.

\subsection{Access structure}

The set of {\it participants}, or agents, who will receive (secret) share 
will
be denoted by $P$. We always assume that $P$ is not empty, and to avoid
certain trivialities, we assume further that is has at least two members.
We allow $P$ to be infinite as well; for the time being
the reader may assume that $P$ is finite. Certain subsets of participants
are expected to recover the secret. The collection of these subsets
is usually denoted by the letter $\mathcal A \subseteq 2^P$, and is called
{\em access structure}. Here $2^P$ denotes the collection of all subsets of
$P$.

\begin{definition}\label{def:access-structure}\rm
$\mathcal A\subseteq 2^P $ is an {\em access structure} if it satisfies the
next two properties:
\begin{enumerate}
\item {\it Monotonicity}: if $A\in\mathcal A$ and $A\subset A'\subseteq P$,
then $A'\in\mathcal A$. Intuitively, if a set is allowed to recover the
secret, then adding further members to this set should not take away this
property.
\item {\it Nontriviality}: there is at least one qualified set 
(and thus $P\in \mathcal A$), furthermore no singleton set is in $\mathcal
A$ (in particular, the empty set is never qualified).
\end{enumerate}
\end{definition}
\noindent
Subsets of $P$ in the access structure $\mathcal A$ are called {\it
qualified}, and subsets not it $\mathcal A$ are {\it unqualified}. As $P$ is
always qualified (as there is at least on qualified set, which is always a
subset of $P$), and no singleton can be qualified, access structures over
$P$ exist if and only if $P$ has at least two elements.

Requiring monotonicity is quite natural: if a group of participants can
recover the secret, then adjoining any further members cannot take away this
ability. The nontriviality means that none of the participants should be
aware of the secret, and also the secret should not be a public knowledge
(which is equivalent to the condition that the empty set is qualified).

Given any collection $\mathcal B$ of subsets of $P$, the smallest monotone
family extending $\mathcal B$ is said to be {\it generated by $\mathcal
B$}. It is easy to check that this is just the collection of supersets of
the elements of $\mathcal B$:
$$
\mathop{\mathrm{gen}}(\mathcal B) \defeq  \{ A\subseteq P \,:\, \mbox{
$B\subseteq A$ for some $B\in\mathcal B$}\,\}.
$$
By the monotonicity property, an access structure is determined uniquely by
any of its generators.
\begin{definition}\label{def:base}\rm
$\mathcal B \subset 2^P$ is a {\em minimal generator} if no proper subcollection of
$\mathcal B$ generates the same collection. Minimal
generators are also called {\em base}.
\end{definition}

\noindent
It is easy to see that $\mathcal B$ is a base if and only if no two elements
in $\mathcal B$ contain each other. Saying otherwise, $\mathcal B$ is a base
if and only if it is a {\em Sperner
system} \cite{sperner}. A collection $\mathcal A$ can have many generators, but
if it has a base then it has unique generator: that base. This base, if
exists, consists 
of the {\it minimal elements} in $\mathcal A$ whose collection is denoted 
by $\mathcal A_0$:
\begin{equation}\label{eq:minimal-qualified}
   \mathcal A_0 = \{ B\in\mathcal A\,:\, \mbox{no proper subset of $B$ is in } 
       \mathcal A\}.
\end{equation}
When specifying an access structure by listing a collection of qualified
subsets, we tacitly assume that $\mathcal A$ is the collection generated 
by the given subsets. It is often desirable (or illuminating) to give the 
minimal qualified subsets only. From (\ref{eq:minimal-qualified}) one can see
that $\mathcal A$ has a base if and only if every qualified subset contains a
minimal qualified subset. This is the case, for example, if every qualified
subset contains a {\it finite} qualified subset, which always happens when
$P$ is finite. When $P$ is infinite the simplest example for an access
structure without base is the family consisting of all 
infinite subsets of $P$. Here the union of any two -- in fact, any finitely 
many -- unqualified subsets is still unqualified.

For any integer $k>1$ the {\it $k$-threshold structure} is the access
structure consisting of subsets of $P$ with at least $k$ elements.
Frequently the number of participants is also mentioned by speaking about an
$(n,k)$-threshold structure; here $n$ is the number of participants in $P$.
As we allow $P$ to be infinite, this latter notation is not always
appropriate. When we speak about a {\it $k$-threshold structure} we always 
assume that $k>1$ is a natural number, that the set $P$ is known, and then 
this phrase denotes the family
$$
        \{ A \subseteq P \,:\, |A| \ge k \} 
$$
generated by the $k$-element subsets of $P$.

A special case of a threshold structure, which generalizes to infinite sets
as well, is the {\it all-or-nothing} structure. In this case there is only
one qualified set, namely all participants are necessary to recover the secret:
$$
      \{P\}.
$$

\subsection{Probabilistic secret sharing scheme}

A {\it secret sharing scheme} is a method to distribute some kind of
information among the participants so that qualified subsets could recover
the secret's value from their shares, while forbidden subset should have no,
or only limited, information on the secret. The following definition captures
the usual notion of secret sharing scheme without specifying exactly what
``information on the secret'' means.

\begin{definition}\label{def:sss-basic}\rm
We say that the secret sharing scheme $\mathcal S$ {\it realizes the access
structure $\mathcal A$}, if
\begin{enumerate}
\item {\it qualified subsets can recover the secret}: for any $A\in \mathcal
A$, the collective shares assigned to members of $A$ determine the secret's 
value;
\item {\it unqualified subsets have no full information}: if $F\notin
\mathcal A$ then the collective shares of $F$ does not determine the
secret's value.
\end{enumerate}
\end{definition}

Several remarks are due. First, no computational issues are considered, thus
rather than requiring any qualified subset of participants be able to {\em
recover} the secret, we rather say that the collection of shares {\em
determine} the secret. This amounts to assuming infinite computational power,
which is a usual assumption in unconditional cryptography. Second, in
several cases we will only require a weaker recoverability condition. As the
secret and shares come from a probability distribution (see Definition
\ref{def:prob.secret.scheme}), we'll be quite
happy if qualified subsets determine the secret {\em with probability 1}
rather than always.

Third, in Definition \ref{def:sss-basic} we deliberately left out what 
{\it ``does not determine''} means, as the exact definition might depend on 
the type of the 
scheme. It might mean ``no information on the secret at all,``
or ``any secret is possible with any collection of the shares'' (but perhaps
with different probabilities),  or
``the collective shares of $F$ does not determine the secret uniquely
with positive probability.'' We will see examples for each of these
possibilities.

\medskip

This paper concentrates exclusively on schemes where every subset is either
qualified or unqualified. It is quite natural to separate these properties,
and consider {\it access} and {\it forbidden} structures with the
possibility that certain subsets are in neither of them. 
Some of our examples generalize to for these more general cases, others do
not. These generalizations, whenever possible, are left to the
interested reader.

\medskip

To define a secret sharing scheme, we must also define the domain
of (possible) secrets, and the domain of (possible) shares for each 
participant. 

\begin{definition}\label{def:sss-domain}\rm
The {\em domain of secrets} is denoted by $X_s$, and the {\em domain of
shares} for the participant $i\in P$ is denoted by $X_i$. We always assume
that none of these sets are empty, and $X_s$ has at least two elements, i.e.
there is indeed a secret to be distributed.
\end{definition}

\noindent
Sometimes, but not necessarily always, some, or all of these domains might
coincide, can be equipped with some (algebraic or geometric) structure. In
classical secret sharing all domains are finite, but we allow these domains
to be infinite sets as well.

\medskip
Secret sharing schemes are usually described by referring to a {\em dealer}
who chooses the secret's value, determines the shares, and distributes
(privately) the shares to the participants. The dealer is assumed to be
honest, following the instructions exactly, and disappearing after her task
is completed without leaking out any information. 
If the dealer is {\em not} assumed to be honest, then participants,
after receiving their shares, engage in a conversation to verify that
indeed they received a consistent set of shares. These schemes go under the
name of {\em verifiable secret sharing} \cite{cramer-vss}, which is not 
considered in this paper.

In our schemes both the secret and the shares are chosen randomly (but not
independently) according to a given joint probability distributions. In such
cases the scheme is determined by the joint distribution of the secret and
of the totality of shares. This approach is a direct generalization of the 
traditional finite secret sharing methods. Using probabilities has also the
advantage that one can easily define what
``no information'' means: the random variable $\xi$ gives no information on
$\eta$ if and only if they are {\it independent}, that is, the probability 
that $\xi$ is in $U$ and $\eta$ is in $V$ is the product of the two separate
probabilities:
$$
\Prob(\xi\in U \mbox{ and } \eta\in V) = \Prob(\xi\in U)\cdot\Prob(\eta\in V).
$$
Or, in other words, the {\em conditional distribution} $\xi|\eta$ (if
exists) is the same as the unconditional distribution of $\xi$.

\medskip

Recall that we denoted by $X_i$ the set of the (potential) shares of 
the participant $i\in P$, and by $X_s$ the set of (possible) secrets 
where $s$ is a dummy element not in $P$. 
\begin{definition}\label{def:prob.secret.scheme}\rm
A {\em probabilistic secret sharing scheme $\mathcal S$} is a
probability distribution on the set $X=\prod_{i\in P\cup\{s\}} X_i$.
Equivalently, it is a collection of random variables $\langle \xi_i: i\in
P\cup\{s\}\rangle$ with some joint distribution so that $\xi_i$ takes values
from $X_i$. The {\it share of
participant $p\in P$} is the value of $\xi_p$, and the secret is the value
of $\xi_s$. 
\end{definition}

\noindent
Using such a scheme, the dealer simply draws the values $\langle \xi_i:i\in
P\cup\{s\}\rangle$ randomly according to
their joint distribution, and then tells the value $\xi_p$ to 
participant $p\in P$, and sets the secret to $\xi_s$. Quite frequently the
same distribution can be reached first determining (or receiving) the secret 
$s=\xi_s$ according to its {\em marginal} distribution, and then drawing the
shares from the conditional distribution $\langle \xi_p: p\in P \rangle| \xi_s
= s$. To do so we need this marginal distribution to exist.

\medskip

According to Definition \ref{def:sss-basic}, 
the secret sharing scheme $\mathcal S$ {\em realizes the access structure 
$\mathcal A$} if members of any qualified set $A\in\mathcal A$ can recover
the secret's value from their
joint shares (at least with probability 1), while unqualified
subsets have, in some sense, limited information only about the value of the
secret. The examples will illustrate this vague definition.

\section{Schemes using uniform distribution}\label{sec:uniform}

The first group of examples use uniform distribution. The simplest one is
the uniform distribution on the interval $[0,1)$, where the measurable
subsets (events) are those $X\subseteq[0,1)$ which have Lebesgue measure.
The {\em probability} that a uniformly distributed random variable $\xi$
is an element of $X\subseteq[0,1)$ is the Lebesgue measure of $X$.

In the sequel we will use $x\pmod 1$ to denote the fractional part of 
the real number $x$.

\subsection{Chor-Kushilevitz scheme}

Let us start with a scheme of Chor and Kushilevitz from \cite{ChorKush}, 
where
both the secret and the shares are uniform random reals from the unit
interval.

\begin{scheme}\label{scheme:ChorKush}{\rm(Chor, Kushilevitz,
\cite{ChorKush})}
Suppose there are $n$ participants, namely $P=\{1,2,\dots,n\}$ where $n$ is a
fixed (finite) natural number. The dealer chooses
$n$ uniform random real numbers $h_1,\dots,h_n$ from $[0,1)$ independently, 
and then she sets the secret to be $(h_1+\cdots+h_n)\pmod 1$,
i.e., the fractional part of the sum.
The {\rm share} of participant $p\in P$ is the number $h_p$.
\end{scheme}

\noindent
It is clear that all participants together can determine the secret using
the same formula what the dealer did. It is also clear that the secret is
uniformly distributed in the unit interval, moreover it is independent of
any $n-1$ of the shares. Thus scheme \ref{scheme:ChorKush} is {\it perfect},
namely unqualified subsets have no information of the secret: the collection
of their shares is {\em independent} of the secret. The scheme
and realizes the all-or-nothing structure $\mathcal A = \{P\}$.

\medskip

The same joint distribution of the secret and the shares can be achieved by
a slightly different procedure. Namely, the dealer could choose the secret and
$n-1$ out of the $n$ shares independently as uniform random reals from
$[0,1)$. Finally she can set the last share from $[0,1)$ so that secret be 
equal to the fractional part of the sum of the shares.

\subsection{A perfect scheme for finitely generated structures}

Let $P$ be arbitrary (maybe infinite), and let $\mathcal A \subset
2^P$ be any access structure {\it generated by finite sets}. That is, every
qualified set (element of $\mathcal A$) should contain a {\it finite} qualified
subset. Using a standard trick due to Ito and al.~from \cite{Ito}, Scheme
\ref{scheme:ChorKush} can be used as a building block to create a perfect
secret sharing scheme realizing $\mathcal A$ which distributes a uniform 
random real from the unit interval $[0,1)$.

\begin{scheme}\label{scheme:ChorKush-general}
Suppose $\mathcal A$ is generated by $\mathcal A_0$ where each element of
$\mathcal A_0$ is finite. Let $s\in[0,1)$ be a uniformly chosen random real,
and for each $A\in \mathcal A_0$ use Scheme \ref{scheme:ChorKush} to
distribute $s$ among the members of $A$ using uniform random $[0,1)$ reals in each
such sub-scheme independently. In this sub-scheme participant $p\in A$ receives the
share $h^A_p\in[0,1)$ (labeled by the minimal qualified subset $A$) such 
that $s$ is the fractional part of the sum $\sum_{p\in A} h^A_p$. 
The total share of participant $p\in P$ will be the tuple $\langle h^A_p:
p\in A$ and $A\in\mathcal A_0\rangle$.
\end{scheme}
\noindent
This is indeed a prefect scheme realizing $\mathcal A$. For the easy part:
if $A\subseteq P$ is qualified, then the secret can be recovered from the
shares members of $A$ received. There is an $A_0\subseteq A$ with $A_0\in
\mathcal A_0$, and members of $A_0$, adding up their shares indexed by 
$A_0$ and taking the fractional part of the sum get the secret.

For the hard part: let $F\subseteq P$ be unqualified (``independent''), we must
show that their total share is independent from the secret $s$. By standard
probability argument \cite{kallenberg} this happens if and only if every
finite subcollection of these values is independent of $s$. So let these
values be $h^{A_1}_{p_1},\dots,h^{A_k}_{p_k}$, where each $p_j\in F$, and
$p_j\in A_j\in \mathcal A_0$. Partition these sub-shares according to their
upper index. Fix $A=A_j$ and let
$h^A_{p_{j_1}},\dots,h^A_{p_{j_\ell}}$ be all the shares in this partition. As
$F$ is unqualified, there is a $p=p_A\in A$ such that the sub-share
$h^A_{p_A}$ is
{\it not} in the partition. Consequently, by the remark following the
description of Scheme \ref{scheme:ChorKush} above,
the shares
$h^A_{p_{j_1}},\dots,h^A_{p_{j_\ell}}$ {\it and the secret} are totally
independent. For different $A\in \mathcal A_0$ the dealer computed the
shares independently, thus any collection of finitely many
sub-shares is indeed independent of the secret.

\subsection{A (not so) perfect 2-threshold scheme}

Let $k\ge 2$ be a (finite) natural number. By instantiating Scheme
\ref{scheme:ChorKush-general} by the generating family $\mathcal A_0 = \{
A\subseteq P \,:\, A$ has exactly $k$ elements$\}$, we see that every 
$k$-threshold structure (independently of whether $P$ is finite or infinite) 
can be realized by
a perfect scheme which distributes a single real number from the unit
interval. As a share, each participant receives as many real numbers as many
minimal ($k$-element) qualified subsets he is in. When $P$ is infinite, this
means infinitely many real numbers. For finite $n$ Shamir's $k$-threshold
scheme \cite{Shamir} distributes a secret so that shares are of the same
size as the secret -- in fact, they are coming from the same algebraic
structure. Generalizing Shamir's scheme for infinitely many participants,
however, raises problematic issues. 

Scheme \ref{scheme:blakley-swanson} below was proposed by Blakley and
Swanson in \cite{blakley-swanson}. It is a geometric version of Shamir's
$2$-threshold scheme \cite{Shamir} for infinitely many participants: any
pair of the participants must be able to determine the secret, but any single
share should be independent from the secret. The idea can be outlined as
follows. The dealer chooses a random line $t$ in the plane. The secret is
the value where $t$ intersects the $y$ axis. Each participant $p$ has a
publicly known non-zero value $x_p$ (the participant's ``label''), and 
$p$'s share is the height where the random 
line intersects the vertical line going through the point $(x_p,0)$. Two
participants can, of course, determine the line $t$, and thus the secret, while
a single participant knows only a single point on $t$, which allows the
secret to be any value whatsoever.

But this is not enough: the secret should be {\it statistically independent} of every
share. Independence can be achieved by using finite geometry, and it also
lets the secret (and all shares) be chosen uniformly. On the Euclidean plane
$\mathbb R^2$ no uniform distribution exists, thus one has to look for other
possibilities. The {\it projective plane} seems to be more promising as it can be
equipped naturally with a finite, uniform measure. Gluing together
diagonally opposite points of the surface of the 3D-sphere gives a
topological equivalent of the projective plane. ``Projective lines''
correspond to main (big) circles and ``projective points'' correspond to
diagonally opposite pairs of points. Mapping diagonally opposite points to
the main circle halfway between them gives the (projective) duality between
points and lines. The uniform distribution on projective points corresponds
to the Lebesgue-measure on measurable symmetric subsets on the surface of
the sphere. By duality, it also gives a uniform distribution on the
projective lines as well. Thus one can choose projective points and lines
``uniformly and randomly.''

\begin{scheme}\label{scheme:blakley-swanson}{\rm(Blakley and Swanson, \cite{blakley-swanson})}
Let $Q$ be a point of the projective plane, and $\ell$ be a line
passing through $Q$, see Figure \ref{fig:projective-plane}. The set of participants is $I-\{\ell\}$ where 
$I$ is the set of all lines through $Q$.
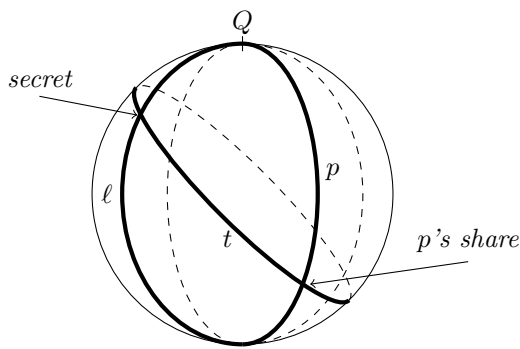
\begin{figure}[htp]
\begin{center}
\begin{tikzpicture}[scale=1.0]
\draw (0,0) circle (2);
\draw[rotate=135,ultra thick](2,0) arc (0:180:2 and 0.4);
\draw[rotate=135,dashed](-2,0) arc(180:360:2 and 0.4);
\draw[ultra thick](0,2) arc(90:270: 1.6 and 2);
\draw[dashed] (0,-2) arc(-90:90:1.6 and 2);
\draw[ultra thick](0,-2) arc(-90:90:1 and 2);
\draw[dashed] (0,2) arc(90:270:1 and 2);
\draw (0,2) node[above] {$Q$};
\draw (0,1.9)--(0,2.1);
\draw (-1.8,0) node{$\ell$};
\draw (-0.2,-0.6) node {$t$};
\draw (1.2,0.3) node {$p$};
\draw[->] (-2.7,1.3) node[above] {\it secret} -- (-1.4,1.05);
\draw[->] (3.0,-0.9) node[above]{\it $p$'s share} -- (0.9,-1.2);
\end{tikzpicture}
\end{center}
\kern -10pt
\caption{$2$-threshold scheme using projective plane}\label{fig:projective-plane}
\end{figure}
The dealer chooses a line $t$ randomly with uniform distribution. With probability $1$ this line
avoids the point $Q$. The secret will be the intersection of $t$ and $\ell$,
and the share of participant $p$ is the intersection of the lines $t$ and $p$.
\end{scheme}

\noindent
Scheme \ref{scheme:blakley-swanson} is a $2$-threshold scheme,
meaning that any two participants can recover the secret, while a single
participant has limited information on the secret. Indeed, any pair of participants
can recover the secret line $t$ as the unique line passing through their
points, thus they can recover the secret (as the intersection of $t$ and
$\ell$) as well.

It is clear that the secret is distributed uniformly on $\ell$, and also
that participant $p$'s share is uniformly distributed on his line as well.
It is also clear that knowing a single share lets every secret possible. One
may be tempted to assume that the {\it conditional distribution} of the
secret given the share of participant $p$ is also {\it uniform}.
\begin{figure}[htb]
\begin{center}
\begin{tikzpicture}[scale=0.51]
\draw (-0.5,0) -- (10.5,0) ; \draw (5,-0.1) node[below] {$\scriptstyle
0$} --(5,12.1);
\draw (0,-0.1) node[below] {$\scriptstyle -\pi/2$} -- (0,12.1);
\draw (10,-0.1) node[below] {$\scriptstyle \pi/2$} -- (10,12.1);
\draw (-1.4,0.25) node {$\scriptstyle d= \pi/18$};
\draw (-1.4,0.7) node {$\scriptstyle d=\pi/9$};
\draw (-1.4,1.35) node {$\scriptstyle d=\pi/4$};
\draw (-1.4,2.05) node {$\scriptstyle d=\pi/2$};
\input gomb.dat
\end{tikzpicture}
\end{center}
\kern -10pt
\caption{Conditional distribution of the secret in Scheme
\ref{scheme:blakley-swanson} for different distances}\label{fig:projective-plane-cond}
\end{figure}
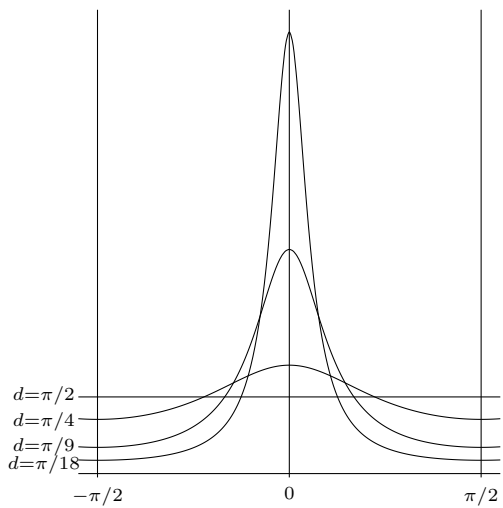
Unfortunately, this is not the case. Fixing the share of $p$ to be some
point $R$, the conditional distribution of the secret line $t$ (now it goes
through $R$) is rotationally symmetric. Thus the conditional distribution of
the secret, that is, the intersection of $t$ and $\ell$ depends (only) on
the distance $d$ between $R$ and $\ell$, and its density function is depicted on
Figure \ref{fig:projective-plane-cond} for different values of $d$.
When the distance is the maximal $d=\pi/2$, then the distribution is uniform
(straight line). The smaller the distance the larger the bump at zero (the
closest point of $\ell$ to $R$). Nevertheless, the minimum is always
positive, thus whatever $p$'s share is, thus the secret can be in any arc of
$\ell$ with positive probability.

\subsection{A perfect 2-threshold scheme}

As noted above, the lack of uniform distribution on the real line prevents a
straightforward application of Shamir's scheme \cite{Shamir} for the case
when the secret is a real number. There are, however, some clever tricks to
overcome this difficulty. Scheme \ref{scheme:uniform-real} is a perfect
$2$-threshold scheme with continuum many participants where the
secret is a uniformly chosen real number from the unit interval.
Unfortunately we were unable to generalize this construction to a
$k$-threshold scheme for any $k$ bigger than 2.

Recall that 
$x\pmod 1$ denotes the fractional part of the real number $x$.

\begin{scheme}\label{scheme:uniform-real}{\rm(Csirmaz, Ligeti, Tardos,
\cite{csirmaz-tardos-ligeti})}
In this scheme the secret $s$ is a uniform real number chosen from the unit
interval $[0,1)$, and participants are indexed by positive real
numbers between $0$ and $1$. The dealer, after choosing the secret $s$ as
described, chooses a uniform random real $r$ from $[0,1)$ independently from
$s$. Participant
with index $p\in(0,1)$ will receive the share $s_p =r+s\cdot p \pmod 1$.
\end{scheme}

\noindent
First note that each participant receives a uniform random number from the
interval $[0,1)$ {\it which is independent of the secret}. This is so as the
``randomization'' value $r$ is independent from the secret (and from the
participant). Thus no participant has any information on the secret
whatsoever, i.e., the scheme is {\it perfect}. Second, if $0< p<q < 1$ are
the indices of two participants, then the fractional part of the difference
of their shares is
\begin{equation}\label{eq:scheme-uniform}
    s\cdot(q-p)~~ \pmod 1 .
\end{equation}
They also know that $s$ is between zero and one, $q-p$ is between $0$ and
$1$, thus $s\cdot(q-p)$ is between $0$ and $1$. By (\ref{eq:scheme-uniform})
they know the fractional part of this product, consequently they know the
exact value as well as it equals to its fractional part. From here
determining the secret is a simple matter of division.

In an intuitive sense this scheme is also {\it ideal}. The shares and the
secret are very much alike, and we did not squeeze extra information into
the shares using some tricky encoding.

\subsection{An ``all-or-nothing'' scheme}

All-or-nothing schemes realize the simplest possible access structure,
namely the one where the only qualified subset consists of all participants:
$\mathcal A = \{P\}$. When $P$ is finite, say has $n$ elements, then Scheme
\ref{scheme:ChorKush} of Chor and Kushilevitz is, in fact, an all-or-nothing
scheme distributing a real number. Thus all-or-nothing schemes exist for
every finite set of participants. Scheme \ref{scheme:weak-ramp}
below is an all-or-nothing one where $P$ has
countably many elements and the secret is a real number.

\begin{scheme}\label{scheme:weak-ramp}{\rm(Csirmaz, Ligeti, Tardos,
\cite{csirmaz-tardos-ligeti})} Let the set of participants be labeled by the
positive integers $\mathbb N^+=\{1,2,\dots\}$. Participant $i\in\mathbb N^+$ 
receives $h_i$, an independent, uniformly distributed real number from
$[0,1)$. Finally the dealer computes the secret as
$$
    s= \sum_{i=1}^{\infty} \frac{h_i}{2^i}.
$$
\end{scheme}

\begin{figure}[htp]
\begin{center}
\begin{tikzpicture}[scale=1.4]
\draw (-0.1,0) -- (5.1,0) ; \draw (2.5,-0.1) node[below] {$\scriptstyle
0.5$} --(2.5,2.1);
\draw (0,-0.1) node[below] {$\scriptstyle 0$} -- (0,0.1);
\draw (5,-0.1) node[below] {$\scriptstyle 1$} -- (5,0.1);
\input distr.dat
\end{tikzpicture}
\end{center}
\caption{Distribution of the secret in Scheme \ref{scheme:weak-ramp}}\label{fig:weak-ramp}
\end{figure}
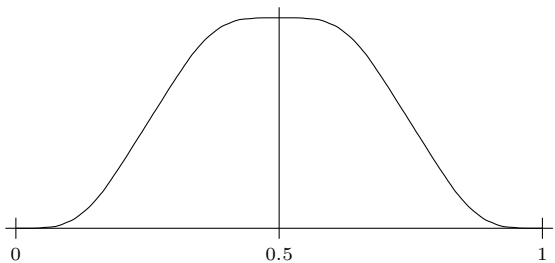

\noindent
The distribution of the secret is depicted on Figure \ref{fig:weak-ramp}.

We claim that Scheme \ref{scheme:weak-ramp} realizes the access structure
$\mathcal A = \{P\}$. First, it is clear that all participants together can
determine the secret: they simply use the same formula what the dealer
did.

Second, suppose participant with label $i\ge 1$ is missing, i.e., only the
shares of $F=P-\{i\}$ are known. As
$i$'s contribution to the secret is a uniform value from an interval of
length $2^{-i}$, members of $F$ only know that $s$ is within a certain
interval (which interval can be computed from their shares), and the secret
has a uniform distribution within that interval. Consequently $F$ can
compute the secret {\em up to an interval of length $2^{-i}$}, and the
conditional distribution of the secret in that interval is uniform. Thus,
in some sense, unqualified subsets have limited information on the
secret. But this ``limited information'' can be quite big: an unqualified
subset can narrow down the value of the secret to a tiny interval. Scheme
\ref{scheme:ell2} realizes the same access structure (namely, $\mathcal
A=\{P\}$), and has the advantage
that unqualified subsets -- based on their shares -- cannot exclude any
secret value whatsoever.

\section{Schemes with Gaussian distribution}\label{sec:gauss}

While there is no uniform distribution on the reals, there is a natural, and
much used one: the {\it Gaussian} or {\it normal} distribution. As usual,
$N(\mu,\sigma^2)$ denotes the normal distribution with expected value $\mu$ and
variance $\sigma^2$, its density function is
$$
    \frac{1}{\sqrt{2\pi\sigma^2}} e^{-\frac{(x-\mu)^2}{2\sigma^2}} .
$$
A {\it standard normal} variable has expected value $\mu=0$ and variance
$\sigma^2=1$. If
$\xi_i$ is normal with parameters $\mu_i$ and $\sigma_i$, and $\xi_1$,
$\dots,\xi_m$ are {\it independent}, then the linear combination $\sum
\lambda_i\xi_i$ is also normally distributed 
with expected
value $\big(\sum \lambda_i\mu_i\big)$ and variance $\big(\sum
\lambda_i^2\sigma_i^2\big)$.

\subsection{A $k$-threshold scheme over the reals}

\begin{scheme}\label{scheme:gauss}
Let $k\ge 2$ be a fixed positive integer. Participants are identified by
non-zero real numbers $\mathbb R - \{0\}$. The dealer sets the share of
participants $x_1=1/k$, $x_2=2/k$, $\dots, x_k=k/k=1$ to be
independent standard normal values. Next, she computes the unique
polynomial $f(x)$ of degree at most $k-1$ for which $f(x_i)$ is the share of
$x_i$ for $i=1,2,\dots,k$. Then she sets the secret as $f(0)$, and
$p\in\mathbb R-\{0\}$ receives the share $f(p)$.
\end{scheme}

\noindent
As a polynomial of degree at most $k-1$ is determined uniquely by values it
takes at $k$ different places, any $k$ participant can compute $f$, and thus
can recover the secret as well. Determining $f$ can be done, for example, by the {\it
Lagrange interpolation method} outlined below.

Suppose we are looking for a polynomial $f$ of degree at most $k-1$ which 
takes the value $y_i$ at $x_i$ for $i=1,\dots,k$. Let $L_i(x)$ be the
following polynomial of degree $k-1$
$$
    L_i(x) = \frac{(x-x_1)\cdots(x-x_{i-1})(x-x_{i+1})\cdots(x-x_k)}
       {(x_i-x_1)\cdots(x_i-x_{i-1})(x_i-x_{i+1})\cdots(x_i-x_k)}
  = \prod_{\scriptstyle j=1\atop \scriptstyle j\ne i}^k \frac{x-x_j}{x_i-x_j},
$$
which takes $1$ at $x_i$ and zero at $x_j$ when $j\ne i$. The polynomial we
are seeking for is
\begin{equation}\label{eq:interpolation}
    f(x) = \sum_{i=1}^k y_i L_i(x),
\end{equation}
as it takes $y_i$ at $x_i$, and has degree at most $k-1$. Equation
(\ref{eq:interpolation}) can also be used to determine the value of $f$ at 
any particular place $x$ without computing the coefficients of $f$ first.

In Scheme \ref{scheme:gauss} the dealer has chosen the values $f(1/k)$,
$f(2/k)$, $\dots,$
$f(k/k)$. Therefore she can compute the share of $p$ using Lagrange
interpolation (\ref{eq:interpolation}) as follows:
$$
    f(p) = \sum_{i=1}^k \bigg( 
      \prod_{\scriptstyle j=1\atop\scriptstyle  j\ne
i}^k\frac{p-j/k}{(i-j)/k}\bigg) f(i/k) = 
      \sum_{i=1}^k \lambda_{p,i} f(i/k),
$$
where the constants $\lambda_{p,i}$ depend only on $p$ (and $i$ and $k$), 
but they do not
depend on the actual values of $f(i/k)$. Thus $f(p)$ is a linear combination of
independent standard normal variables, consequently it also follows a Gaussian
distribution with expected value $0$ and variance
$$
  \sigma_p^2 =  \sum_{i=1}^k \lambda_{p,i}^2 =
    \sum_{i=1}^k \prod_{j\ne i} \frac{(kp-j)^2}{(i-j)^2}.
$$
In particular, the secret, namely the value of $f$ at $0$, has expected
value $0$ and variance
\begin{equation}\label{eq:gauss-secret-variance}
  \sigma_0^2 = \sum_{i=1}^k \prod_{j\ne i} \frac{j^2}{(i-j)^2} =
   \sum_{i=1}^k {k\choose i}^2 = {2k\choose k}-1 \approx 
    \frac{2^{2k}}{\sqrt{\pi k}}.
\end{equation}
The {\it joint distribution} of the shares of $k-1$ participants $p_1,
\dots,p_{k-1}$ is a $(k-1)$-dimensional Gaussian distribution; these shares,
however, are not necessarily independent. Nevertheless, the {\it conditional
distribution} of the secret, given the shares $f(p_1)=h_1, \dots,
f(p_{k-1})=h_{k-1}$ is again Gaussian. Indeed, the probability that
$f(p_j)=h_j$ and $f(0)=s$ (namely the value of the density function at this
point) can be
computed as follows. Using (\ref{eq:interpolation}) we can compute $f(1/k),
\dots,f(k/k)$ from the values $h_j$ and $s$ as
$$
    f(\ell/k) = s \big(\prod_{j=1}^{k-1} \frac{\ell/k-p_j}{0-p_j}\big) +
    \big( \sum_{i=1}^{k-1} h_{i} \frac{\ell/k-0}{p_i-0}\prod_{j\ne
    i}\frac{\ell/k-p_j}{p_i-p_j}\big)
    = s\cdot a_\ell-b_\ell,
$$
here $\ell=1,\dots,k$, and the constants $a_\ell$ and $b_\ell$ do not depend on $s$. Thus
\begin{eqnarray}
&&\hskip -20pt \Prob\big(f(0)=s, f(p_1)=h_1,\dots,f(p_{k-1})=h_{k-1}\big) \,= \nonumber\\[3pt]
&=& \Prob\big(f(1/k)=sa_1-b_1,\dots,f(k/k)=sa_k-b_k\big) \,=\nonumber\\
&=& \frac1{(2\pi)^{k/2}} \exp \big( {\textstyle-\frac12} ((sa_1-b_1)^2+\cdots 
               + (sa_k-b_k)^2)\big) \,=\nonumber\\
&=& C\cdot\exp\big({\textstyle-\frac12} (As-B)^2\big),\label{eq:conditional}
\end{eqnarray}
where
\begin{eqnarray*}
A^2 &=& a_1^2 + \cdots a_k^2 =\sum_{\ell=1}^k  \frac{(\ell/k-p_1)^2
\cdots(\ell/k-p_{k-1})^2}
    {p_1^2 \cdots p_{k-1}^2},\\[5pt]
AB &=& a_1b_1 + \cdots a_kb_k,
\end{eqnarray*}
and $C$ is again some constant not depending on $s$. According to
(\ref{eq:conditional}), the conditional distribution of the secret is indeed
normal with expected value $B/A$ and variance $A^{-2}$. The variance depends
only on who the participants $p_1, \dots, p_{k-1}$ are, but does not depend
on the value of their shares. If one of the participants is quite close to
zero, then we expect his share to be close to the secret as well, and this 
intuition
is justified by the small value of the variance $A^{-2}$. Also, if all
$p_\ell$ are really small, then in the formulas above we can replace $\ell/k-p_j$ by
$\ell/k$, and the expected value of the conditional distribution of the secret
can be approximated as
$$
\frac{B}{A} = \frac{AB}{A^2}\approx 
\sum_{i=1}^{k-1}h_i\prod_{j\ne i} \frac{0-p_j}{p_i-p_j},
$$
which is the value at zero of the smallest degree polynomial going through
the shared values of these $k-1$ participants, while the variance is
$$
   A^{-2} \approx \frac{p_1^2\cdots p_{k-1}^2}{\sum_{\ell=1}^k (\ell/k)^{2k-2}}
   \approx p_1^2\cdots p_{k-1}^2.
$$
On the other hand, if all participants are well above $1$, i.e., $p_j\gg 1$,
then we can bound the conditional variance $A^{-2}$ from below. As $\ell\le k$,
$$
\frac{(\ell/k-p_j)^2}{p_j^2} = \left(1- \frac\ell{kp_j}\right)^2 < 1,
$$
therefore 
\begin{equation}\label{eq:gauss-variance}
A^{-2}=\left(\sum_{\ell=1}^k  \frac{(\ell/k-p_1)^2
\cdots(\ell/k-p_{k-1})^2} {p_1^2 \cdots p_{k-1}^2} \right)^{-1}
> \frac1k.
\end{equation}
The variance is asymptotically $1/k$ as the values of $p_j$ tend to
infinity; and the expected value is
$$
\frac{AB}{A^2}\approx \sum_{i=1}^{k-1} h_i\frac{k}{2p_i}\prod_{j\ne i}
     \frac{0-p_j}{p_i-p_j},
$$
once again the value at zero of the smallest degree polynomial going through
the ``corrected'' values $\tilde h_i = h_ik/(2p_i)$. Consequently this
scheme is far away from being ``perfect,'' which would require that the
conditional distribution of the secret be the same as the unconditional one.
Nevertheless, the scheme offers quite good security guarantees, namely 
any $k-1$ participants, whatever their shares are, cannot exclude any interval, or in
fact any set with positive Lebesgue measure, from $\mathbb R$ as the 
possible value of the secret.

\subsection{An almost perfect threshold scheme}

\begin{figure}[htb]
\begin{center}
\begin{tikzpicture}[scale=1.0]
\draw (-0.25,0) -- (5.25,0) ; \draw (5,-0.1) node[below] {$1$} --(5,2.8);
\draw (0,-0.1) node[below] {$0$} -- (0,2.8);
\draw (-0.9,1.0) node {$\sigma=3$};
\draw (-0.8,1.6) node {$\sigma=1.5$};
\draw (-0.9,2.3) node {$\sigma=1$};
\input ajtai.dat
\end{tikzpicture}
\end{center}\vspace{-10pt}
\caption{Distribution of the fractional part of a normal variable}\label{fig:ajtai-dwork}
\end{figure}
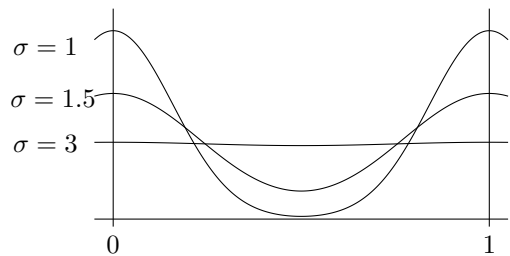

Using a trick inspired by the Ajtai-Dwork cryptosystem \cite{regev}, we can
make the previous threshold Scheme \ref{scheme:gauss} ``almost perfect.'' In 
designing their cryptosystem, Ajtai and Dwork used the fact
that the fractional part of a normal variable with large enough variance is
almost uniformly distributed. The density function of the distribution of 
the fractional part of a normal variable with variance $\sigma^2$
can be computed using the Poisson summation formula
\begin{equation}\label{eq:Poisson}
\frac1{\sqrt{2\pi\sigma^2}}\sum_{k=-\infty}^{+\infty} 
  e^{-\frac{(x-k)^2}{2\sigma^2}} =
\sum_{\ell=-\infty}^{+\infty} \cos(2\pi x \ell) e^{-2\pi^2\sigma^2\ell^2} ,
~~~ 0\le x < 1.
\end{equation}
On figure \ref{fig:ajtai-dwork} this density function is depicted for three 
different values of $\sigma$.
For each $\sigma$ it takes the smallest value at
$1/2$, and the largest value at $0$ (and $1$). The larger the variance the
closer the distribution to the uniform one. In fact, as formula 
(\ref{eq:Poisson}) indicates, the deviation from the
uniform distribution is of order $\exp(-2\pi^2\sigma^2)$, thus choosing, e.g., 
$\sigma=10$ the distribution is practically indistinguishable from
the uniform one. 

The trick is that instead of using a normally distributed random 
variable $\xi\in N(m,\sigma^2)$,
we will
use its fractional part $\eta = \xi\pmod 1$. The density function of
$\eta$ differs from the density function of the uniform distribution by
about $\exp(-2\pi^2\sigma^2)$. 
With this trick in mind we make the following changes to Scheme 
\ref{scheme:gauss}. First, the
secret will not be the value of the interpolating polynomial $f$ at zero, 
rather
the fractional part of it. If $f(0)$ has large enough 
variance, then its
fractional part is almost uniform in the unit interval. Second, we
restrict participants to values larger than $1$ to make sure that the
conditional distribution of the secret has variance separated from zero.
Third, to make sure that $f(0)$ has variance large enough, $f(1/k),\dots,f(k/k)$
will be chosen independently from $N(0,\sigma^2)$ with some large enough 
$\sigma$.

\begin{scheme}\label{scheme:ajtai-dwork}
Let $k\ge 2$ be a fixed positive integer, $\lambda>1$ be a ``security''
parameter, and set $\sigma=\lambda\cdot\sqrt k$.  Participants are labeled by real
numbers above $1$. The dealer chooses $k$ normally distributed random reals
$\xi_i$ from $N(0,\sigma^2)$ independently, and determines the smallest degree
polynomial $f$  with $f(i/k)=\xi_i$. Then she sets the secret to be the
fractional part of $f(0)$, and gives participant $p > 1$ the share
$f(p)$.
\end{scheme}

\noindent
As with Scheme \ref{scheme:gauss} any subset of $k$ participants can recover
the polynomial $f$, thus the secret as well. 
According to equation (\ref{eq:gauss-secret-variance}), $f(0)$ has variance
$$
   \sigma^2 \cdot \sum_{i=1}^k {k\choose i}^2  > \lambda^2\cdot k\cdot 2^k ,
$$
and from (\ref{eq:gauss-variance}) it follows that the variance of the {\it 
conditional distribution} of $f(0)$ given $k-1$ shares is bigger than
$$
   \sigma^2 \cdot \frac1k = \lambda^2 .
$$
Consequently the density function of the fractional part of $f(0)$, i.e.,
that of the secret, differs from the uniform one by less than
$2\exp(-2\pi^2\lambda^2)$
both in the unconditional and in the conditional case. Therefore 
these density functions differ from each other by less than 
$4\exp(-2\pi^2\lambda^2)$ {\it everywhere}.
Choosing, e.g.,  $\lambda=10$, the two distributions are practically indistinguishable,
unqualified subsets have (uniformly) arbitrary small information on 
the secret.

One could slightly twist Scheme \ref{scheme:ajtai-dwork} to realize the {\em
infinite} subsets of participants indexed by positive real numbers above 1.
Shares they receive will form a graph of a polynomial, but in this case the
degree of the polynomial has no a priori bound. As usual, the secret
is the value of the polynomial at zero (or, rather, its fractional part). 
Infinitely many points determine the
polynomial -- and the secret -- unambiguously, while given finitely many
shares only, the ratio of the unconditional and conditional probability of
the secret is bounded.

\begin{scheme}\label{scheme:bounded2}
Fix the ``security parameter'' $\lambda>1$. Participants are labeled by real
numbers above $1$. The dealer chooses the integer $k\ge 2$ randomly so that
$k$ is chosen with probability $2^{-k+1}$, then she executes Scheme
\ref{scheme:ajtai-dwork} with $k$ and $\lambda$.
\end{scheme}

\noindent
Clearly both the secret and the shares are random variables with some
well-determined distribution which depends on $\lambda$. The secret is 
from the unit interval with a distribution quite close to the uniform one.
Any infinite set of participants can determine the polynomial $f$ whatever
its degree is, thus they can determine the secret as well. 

With probability $2^{-k+1}$ the polynomial will be of degree $k-1$, thus $k$
participants can determine the secret uniquely with probability
$1-2^{-k+1}$.  With probability $2^{-k+1}$, however, $f$ has degree at least
$k$, and the conditional distribution of the secret is quite close to the
uniform one (say bigger than $1/2$ everywhere). Consequently, no matter what
the shares of these $k$ participants are, the density function of the
conditional distribution of the secret given these shares is at least
$2^{-k}$ over the whole unit interval. Consequently they have at least that
much uncertainty about the value of the secret.

\subsection{Gaussian ``all-or-nothing'' scheme}

Scheme \ref{scheme:weak-ramp} realized an infinite ``all-or-nothing'' access
structure: all participants were necessary to recover the secret exactly.
When the $i$'th participant's share was missing, others could narrow down the
secret to an interval of length $2^{-i}$ only. Here we present an alternate
scheme realizing the same structure, based on a Gaussian probability
distribution on $\ell_2$ in which unqualified subsets cannot exclude any
positive set as the possible value of the secret.

Suppose $\langle\xi_i\in N(0,\sigma_i^2)\,:\, i\in\mathbb N^+\rangle$ is 
a sequence of independent normal variables. This defines a (Gaussian) 
probability measure on the infinite sequences: to choose such a sequence
``randomly,'' simply take a realization of the sequence 
$\langle \xi_i: i\in\mathbb N^+\rangle$.
If, in addition, $\sigma^2 =\sum_i \sigma_i^2 < \infty$, then, with
probability 1, the sum $\sum \xi_i$ converges, and has normal 
distribution $N(0,\sigma^2)$. Fix such $\sigma_i$ values. Shares will be
generated according to the above measure, and the secret will be the sum
of the distributed vector. 
\begin{scheme}\label{scheme:ell2}
Fix the values $\sigma_i$ such that $\sum\sigma_i^2<\infty$.
Participants are labeled by the set $\mathbb N^+$ of positive integers.
The share $h_i$ of $i\in\mathbb N^+$ is a normally distributed real number 
from $N(0,\sigma_i^2)$ where shares are chosen 
independently. Finally the dealer computes the secret as $s=\sum_i h_i$.
\end{scheme}

\noindent
As remarked above, with this choice of the variances, the sum $\sum_i h_i$ is
convergent with probability $1$, thus this is a correct scheme.
Evidently, all participants together can determine the secret by
adding up their shares. On the other hand, participant $i$ contributes a
normally distributed value to the secret {\em which is independent of all
other shares}, thus leaving him out the others
cannot exclude any set of reals with positive measure as the secret's value.
Indeed: given the shares of a maximal unqualified set
$F=\mathbb N^+-\{i\}$, the {\em conditional distribution} of $s$ given all these
shares is $N(m,\sigma_i^2)$, where $m=\sum_{j\ne i} h_j$, the sum of the
shares of the members of $F$.

\medskip

Using the Ajtai-Dwork trick, namely taking the secret to be the fractional
part of the value to be recovered, this scheme can also be turned 
into a scheme with stronger ``hiding'' property. Namely, given the shares of
an unqualified set $F$, the density function of the conditional distribution
of the secret is not only positive wherever the unconditional secret
distribution has positive density, but the two distributions are only a
``constant apart:''
\begin{equation}\label{eq:up-to-constant}
  c\cdot(\mbox{unconditional density}) \le
    (\mbox{conditional density}) \le \frac1c\cdot(\mbox{unconditional
density})
\end{equation}
for some positive constant $c=c_F$ depending on the unqualified set $F$.

\medskip

We remark that in \cite{infinite-secret-theory} it is shown that infinite
``all-or-nothing'' structures cannot be realized by perfect schemes, thus
this improved scheme provides the strongest possible security guarantee for 
this particular access structure.

\section{Stochastic processes}\label{sec:stochastic}

Schemes in this section are based on the heavy weight champion of
probability theory, the {\it Wiener process}. A Wiener process $W(t)$ is a
randomly generated function defined on the interval $[0,1]$ which is the
limiting process of the scaled sum of infinitely many independent
identically distributed random variables. Among its several properties, we
will use the following ones.
\begin{enumerate}
\item $W(0)=0$ and for $0<t\le 1$, $W(t)$ is normally distributed with
expected value $0$ and variation $t$;
\item $W(t)-W(s)$ is normal with expected value $0$ and variance $|t-s|$;
\item $W(t)$ has {\it independent increments}, that is when
$t_1<t_2<s_1<s_2$ then $W(t_2)-W(t_1)$ and $W(s_2)-W(s_1)$ are independent;
\item the expected value of $W(t)W(s)$ is $\min(t,s)$;
\item $W(t)$ is continuous, and has unlimited variance (with
probability $1$);
\item the integral $\int_0^t W(s)\,ds$ is normally distributed with expected
value $0$ and variation $t^3/3$.
\end{enumerate} 
All but the last properties are standard ones, see, e.g., \cite{kallenberg}.
To see why the last property holds, we first remark that $W(s)$ has expected value $0$, thus
by Fubini's theorem the integral has expected value zero as well. The
variance can be computed as follows:
$$\begin{array}{l}
\displaystyle E \big[\big(\!\!\int_0^t W(s)\,ds\big)^2\big] = 
E \big[\int_0^t\!\! \int_0^t W(s)W(s')\,ds\,ds'\big] = {}\\[10pt]
\displaystyle{}=\int_0^t\!\!\int_0^t E\big[W(s)W(s')\big]\,ds\,ds' = 
\int_0^t\!\!\int_0^t \min(s,s')\,ds\,ds' = t^3/3,
\end{array}
$$
as was claimed. 

\subsection{Dense subsets}

\begin{scheme}\label{scheme:dense}
The participants are labeled by real numbers from the unit interval
$[0,1]$. The dealer chooses a Wiener process $W(t)$, and tells participant
$p\in[0,1]$ the share $W(p)$. Finally, she sets the secret to be
$\int_0^1W(t)\,dt$.
\end{scheme}
\noindent
This scheme realizes the {\it dense subsets of the unit interval}. Indeed,
$W(t)$ is continuous (with probability 1), thus knowing its value on a dense subset determines
the function everywhere, and then the participants can integrate the
function to determine the secret. If $F\subseteq[0,1]$ is {\it not} dense,
then there is a subinterval $[a,b]$ disjoint from $F$, we may assume that
$F$ is just the complement of this interval. Now the secret can be written
as the sum
$$
   \int_I W(t)\,dt = \int_{[a,b]} W(t)\,dt + \int_{I-[a,b]} W(t)\,dt
$$
Both integrals on the right hand side are normally distributed with variance
$(b-a)^3/3$ and $1/3-(b-a)^3/3$ respectively. The independent increments
property of the Wiener process tells us that these summands are independent,
thus members of $F$ know the secret only up to a normally distributed value
with positive variance which is independent of their shares.

\smallskip

Scheme \ref{scheme:dense} can be restricted to any dense subset of $[0,1]$,
in particular to the set of dyadic rationals in $(0,1)$, which, in turn, can
be identified with the nodes of the infinite binary tree $T$. A set $D$ of
the nodes is {\it dense in $T$} if every node has an extension in $D$, or
said otherwise, no spanned subtree of $T$ avoids $D$.

\begin{scheme}\label{scheme:countable-dense}
Participants are nodes of the infinite binary tree $T$ labeled by finite
$\{-1,+1\}$ sequences, including the empty sequence for the root. This
scheme realizes the dense subsets of $T$. The dealer chooses a Wiener
process $W(t)$, and sets the secret to be $\int_0^1 W(t)\,dt$. Participant
with label $\langle \eps_1,\eps_2,\dots,\eps_k\rangle$ where $\eps_i=\pm1$
receives the value of $W(t)$ at
$$
t=\frac12 + \frac12 \sum_{i=1}^k \frac{\eps_i}{2^i} .
$$
\end{scheme}

\noindent
Both the secret and all shares are normally distributed values. Dense
subsets of the participants can recover the process $W(t)$, thus its
integral as well. Subsets missing a whole subtree miss a whole interval from
$[0,1]$, thus have limited information on the secret.

\subsection{Convergent sequences}

Rather than setting the secret to be the integral of the Wiener process, we
could set the secret to be simply $W(1)$. If participants receive $W(t)$
for other values of $t$, then a subset $A\subseteq[0,1)$ can recover the 
secret if and only if $1$ is
in the closure of $A$. Indeed, $W(t)$ is continuous at $a$. Furthermore,
if $A$ is separated from $1$,
i.e., there is a whole interval between $A$ and $1$, then by the independent
increments property, $W(1)$ has an increment (a normally distributed value)
with variance equal to the length of that interval, which is {\it
independent} of the values known to members of $A$. Consequently $A$ has
only a limited information on the secret, and secret's value can be in any
interval with positive probability. Thus Scheme \ref{scheme:limit} below
realizes those subsets of $P$ which have $1$ in their closure:

\begin{scheme}\label{scheme:limit}
The set of participants is $P\subseteq [0,1)$ such that $1$ is in the
closure of $P$. The dealer chooses a Wiener process $W(t)$, sets the secret
to be $W(1)$ and tells participant $p\in P$ the value $W(p)$.
\end{scheme}
\noindent
If we choose $P=\{ 1-1/i: i\in\mathbb N^+\}$ then qualified subsets are
exactly the infinite subsets of $P$. Thus we got a scheme realizing
all infinite subsets of a countable set, see also \ref{scheme:stat1}.

\medskip

Schemes 
in this section 
can also be turned into a stronger scheme with ``up to a constant hiding''
property (\ref{eq:up-to-constant}) by setting the secret to be its 
fractional part.

\section{Statistical methods}\label{sec:statistics}
Recovering the secret can be considered a statistical problem: participants
have values which are ``measurements'' and the are interested in an estimate
on the ``unknown parameter'' $s$. The simplest case is when shares are
``measurements'' of the secret up to an error term with standard normal
distribution.

\begin{scheme}\label{scheme:stat1}
The dealer chooses the secret $s\in N(0,1)$. The share of participant $i\in
P$ is
$s+\xi_i$ where the hiding value $\xi_i\in N(0,1)$ is chosen independently
from all other values.
\end{scheme}

\noindent
When $n$ participants put together their values, they have $n$ independent
measurements of the secret. In this case the best statistics, namely the
best estimate of the unknown value $s$ is the average of their values, see
\cite{kholevo:2001}. The {\em conditional distribution} of the secret
assuming their shares $h_i$ is normal with expected value $\bar h$ (sample
mean), and variance $1/n$. Thus finite subsets are unqualified: they have
limited information on the secret.

Infinite subsets, however, can determine $s$ with probability $1$: their
sample mean has expected value equal to the secret, and has variance zero,
thus is must be equal (with probability 1) to the secret.

Consequently, Scheme \ref{scheme:stat1} realizes all infinite subsets of any
(infinite) set $P$ of participants, with the same security guarantees as
did Scheme \ref{scheme:bounded2} or Scheme \ref{scheme:limit}. Those
schemes, however, worked only when $P$ was countably infinite.

\bigskip
In our second scheme each participant $i\in P$ has a publicly known {\em
obfuscating value $1\le r_i < 10$}.

\begin{scheme}\label{scheme:stat2}
The dealer chooses the secret $s\in N(0,1)$, and the obfuscating value
$\eta\in N(0,1)$. The share of participant $i\in P$ is $(s+r_i\cdot\eta +
\xi_i)$ with $\xi_i\in N(0,1)$, where all random variables ($s$, $\eta$, and
$\xi_i$) are chosen independently.
\end{scheme}

\noindent
Suppose $A\subseteq P$ is an infinite subset of participants whose
obfuscating values have the only limit point $r$. Then the sample mean of
their shares will be $s+r\cdot\eta$ (as the sample mean is normally
distributed with this expected value and zero variance). Consequently a
subset $A\subseteq P$ can determine the secret if the set of their 
obfuscating values has at least two different limit points $r$ and $t$: they
can find out the numbers $r$, $t$, $a=(s+r\cdot\eta)$ and $b=(s+t\cdot\eta)$, 
and then compute the secret as $(t\cdot a - r\cdot b)/(t-r)$.

It is also clear that if $A\subseteq P$ is {\em finite}, then the
conditional distribution of the secret, given the shares of $A$ is normal
with positive variance (in fact, the variance is at least $1/n$ when $A$ has
$n$ elements), thus finite subsets are {\em unqualified}.

The question is when $A\subseteq P$ is infinite and their obfuscating numbers
have a single limit value, when will $A$ be qualified? To answer this
question, let us look at the conditional distribution of the secret given
$n$ shares $h_1, \dots, h_n$ with obfuscating values $r_1,\dots,r_n$. As
$\eta$ and $\xi_i$ are independent standard normal,
$$
  \Prob(s=x,\, \eta=y,\, s+r_i\eta+\xi_i=h_i) =
   \frac1{(\sqrt{2\pi})^{n+2}}\, e^{-x^2/2}\,e^{-y^2/2}\,\prod_{i=1}^n
        e^{-(h_i-x-r_i\cdot y)^2/2} .
$$
Thus the conditional density function of $s$ given the shares
$h_1,\dots,h_n$ is the integral of this function by $y$. Concentrating on
the exponent only,
\begin{eqnarray*}
 &&  x^2+y^2+\sum_{i=1}^n(h_i-x-r_i\cdot y)^2 = {}\\
 &=& x^2+(1+\tsum r_i^2)y^2 + 2y\tsum r_i(h_i-x) + \tsum(h_i-x)^2 =
 {}\\[5pt]
 &=& (1+\tsum r_i^2)(y+A)^2 - (1+\tsum r_i^2)A^2 + x^2 + \tsum(h_i-x)^2, 
\end{eqnarray*}
where
$$
  A = \frac{\strut\tsum r_i(h_i-x)}{\rule{0pt}{3ex} 1+\tsum r_i^2}.
$$
When integrating, the value of the first term becomes a constant (its value
does not depend on $A$, thus on $x$), therefore the exponent of the density
function of the conditional distribution is
\begin{eqnarray*}
&& - (1+\tsum r_i^2)A^2 + x^2 + \tsum(h_i-x)^2 = {}\\[5pt]
&=& x^2\left(n+1-\frac{\strut \big(\sum r_i\big)^2}{\rule{0pt}{3ex} 1+\sum r_i^2}\right) + Bx + C
={}\\
&=& x^2\left(1+\frac{\strut 1+\sum(r_i-\bar r)^2}{\rule{0pt}{3ex}
  \bar r^2 + \frac1n\big(1+\sum(r_i-\bar r)^2\big) }\right) + Bx + C,
\end{eqnarray*}
where $\bar r$ is the average (mean) of the $r_i$'s, and $B$, $C$ are
expressions which do not depend on $x$. Consequently, the conditional
distribution of the secret, given the shares $h_i$, is normally distributed
with variance $\sigma^2$, where
\begin{equation}\label{eq:stat2}
\frac1{\sigma^2} = 1+\frac{\strut 1+\sum(r_i-\bar r)^2}{\rule{0pt}{3ex}
  \bar r^2 + \frac1n\big(1+\sum(r_i-\bar r)^2\big) }.
\end{equation}
If $A\subseteq P$ is infinite and the obfuscating numbers of $A$ has the single
limit $r$, then, again, the conditional distribution of the secret is
normally distributed, and the variance can be computed from equation
(\ref{eq:stat2}), but in this case the average $\bar r$ coincides with the
limit $r$. The subset $A$ can determine the secret if and only if the
variance $\sigma^2$ is zero. And this happens, if and only if
$$
    \sum_i (r_i-r)^2 = \infty .
$$
(Note that, by assumption, $r\ge 1$.) Thus $A\subseteq P$ with a single
limit $r$ is a) unqualified if $\sum (r_i-r)^2 < \infty$, and in this case the
conditional distribution of the secret is normal with a positive variance;
and b) $A$ is qualified if the sum $\sum (r_i-r)^2$ diverges.

\smallskip
\hfill$*$\hfill
\smallskip

Let $P$ be the set of lattice points in the positive quadrant, that is the
set of point with positive integer coordinates. A {\em ray} is a
half line starting from the origin. We consider only those rays which are in
the positive quadrant. The {\em angle} or {\em argument} of the lattice point
$p\in P$ is the angle between the ray going through $p$ and the
positive $x$ axis. All these angles are between $0$ and $\pi/2$.
A {\em strip of width $d$} is a subset
$S\subseteq P$ for which there is a ray lying in the first quadrant -- the
{\em direction} of the strip -- so that a lattice point is in $S$ if and
only if it is at distance at most $d/2$ from the ray. Strips with different
directions have only finitely many lattice points in common. Considering all
strips of width, say, 10, is the standard way to construct continuum many
infinite subsets of a countable set ($P$ in this case) so that any two of 
them has finite intersection, see \cite{kunen1983set}.

\begin{scheme}\label{scheme:almost-disjoint}
The set $P$ of participants in this scheme is the set of lattice points of
the positive quadrant. The dealer chooses the secret $s\in N(0,1)$, and the
obfuscating value $\eta\in N(0,1)$. The share of participant $p$  with
angle $\phi$ is $(s+(1+\phi)\eta+\xi_p)$, where
$\xi_p\in N(0,1)$, and, as before, all random variables are chosen
independently.
\end{scheme}

\noindent
We claim first that every strip is unqualified. Indeed, let $\phi$ be the
the angle of the direction of the strip $S$, and let $d>5$ be its width. For each
natural number $j>0$, there are at most $100d$ lattice points within the strip 
with distance between $j$ and $j+1$ from the origin. Each of these lattice
points have angles between $\phi-d/j$ and $\phi+d/j$. Consequently the
obfuscating numbers of the lattice points in this strip have a single 
limit value, namely $r=(1+\phi)$, and
$$
   \sum_{p\in S} (r_p-r)^2 < \sum_{j>0} (100d)\cdot(d/j)^2 =
      100d^3 \sum_{j>0} \frac1{j^2} < +\infty.
$$
As has been shown previously, this means that $S$ is unqualified.

On the other hand, the union of any two strips with different directions is
qualified. Indeed, in this case the obfuscating numbers of the subset have 
two different limit points, which is a sufficient condition for a subset to
be qualified.

Thus Scheme \ref{scheme:almost-disjoint} realizes an access structure over a
countable set of participants in which 
there are continuum many unqualified
subsets such that the union of any two of them is qualified.

\section{Hilbert space program}\label{sec:hilbert}

Span programs, introduced by Karchmer and Wigderson in
\cite{spanprogram}, provide a general framework for defining and
investigating (finite) linear secret sharing schemes. Instead of random
variables span programs are defined on (finite dimensional) vector spaces.
Let $\mathbb V$ be such a vector space, and fix $u\in\mathbb V$ t as the 
{\em goal vector}.
Every participant $p\in P$ is assigned a linear subspace $L_p\subseteq
\mathbb V$. A
collection of participants is qualified, if the linear span of their
subspaces contain the goal vector, and unqualified otherwise.

If $\mathbb V$ is finite, that is the underlying field $\mathbb F$ is finite,
then there is a
natural way to convert a span program into a secret sharing scheme. Choose
$r\in\mathbb V$ randomly with uniform distribution. The secret will be the
inner product $r\cdot u$. If $L_p$ is $k$-dimensional and is
spanned by $x_1,\dots,x_k$, then $p$'s share is the $k$-tuple 
$\langle r\cdot x_1$, $\dots,$ $r\cdot x_k\rangle$. In this way qualified
subsets can recover the secret, and the shares of an unqualified subset give
no information on the secret: the $r\cdot u$ inner product can be any
element of $\mathbb F$ with the same probability.

A generalization of this notion to infinitely many participants is the {\em
Hilbert space program}

\begin{definition}\label{def:hilbert-space-program}
Let $H$ be a (real) Hilbert space, and fix $u\in H$ as the goal. In a {\em
Hilbert space program} each participant $p\in P$ is assigned a subspace
$L_p\subseteq H$. A collection of participants is qualified, if the goal
vector is in the closure of the linear span of their subspaces, and
unqualified otherwise.
\end{definition}

Just as in the case of span programs, every Hilbert space program can be
turned into a secret sharing scheme. The idea is that make $H$ a
Gaussian space \cite{janson1997gaussian}. Let $B$ be an orthonormal basis of
the Hilbert space $H$,
and choose the standard normal variables $\xi_e\in N(0,1)$ for each $e\in B$
independently. Every element $x\in H$ can be written uniquely as 
$x =\sum\{\lambda_e(x) e \,:\,e\in B\}$, where $\lambda_e(x)$ is just the
inner product of $x$ and $e$. To the element $x$ associate the
random variable $\xi_x = \sum\{\lambda_e(x)\xi_e\}$. This will be a 
centered Gaussian random variable with variance $\|x\|^2$. 

Set the secret to be the value of $\xi_u$, the random variable assigned 
to the goal $u\in H$. If
participant $p$ got the subspace $L_p$, then let $B_p\subseteq L_p$ a base
in it, and $p$'s share will be the values of $\xi_b$ for $b\in B_p$. 
It is clear that qualified subsets can determine the secret. Indeed, they
can use the linear combination which produces the goal from their
vectors.
Next suppose $A\subseteq P$ is
unqualified. Let $L\subseteq H$ be the closure of the subspace spanned by
the family $\{L_p \,:\,p\in A\}$. Let $v$ be the orthogonal projection of $u$ 
into
$L$, and let $w=u-v$. As the goal is not in $L$, $w\not=0$. 
Then $v$ and $w$ are orthogonal, and $u=v+w$. This
means that the secret $\xi_u$ is the sum of $\xi_v$ and $\xi_w$, and that $\xi_v$ and
$\xi_w$ are uncorrelated, thus independent. But $w$ is also orthogonal 
to the whole
subspace $L$, thus $\xi_w$ is independent of all shares in $A$.
Consequently the conditional distribution of the secret, given all shares of
$A$, is normal with variance $\|w\|^2$, that is $A$ has that much uncertainty
about the secret's value.

\bigskip
Schemes \ref{scheme:gauss}, \ref{scheme:ell2}, \ref{scheme:dense},
\ref{scheme:countable-dense}, \ref{scheme:limit}, \ref{scheme:stat1}, and
\ref{scheme:stat2} are all instances of this general construction. For
example, in Scheme \ref{scheme:stat2} the Hilbert space is separable, i.e.,
it has a countable orthonormal base.
Each participant gets a
one-dimensional subspace. The coordinates of the goal and the vectors which
span these subspaces are
$$
\begin{array}{ccccccccc}
\mbox{goal:}~~ & 1 & 0   & 0 & 0 & 0 & 0 & 0 & \dots \\[3pt]
\hline
\mbox{shares:\rule{0pt}{3ex}}  & 1 & r_1 & 1 & 0 & 0 & 0 & 0 & \dots \\
            & 1 & r_2 & 0 & 1 & 0 & 0 & 0 & \dots \\
            & 1 & r_3 & 0 & 0 & 1 & 0 & 0 & \dots \\
            & 1 & r_4 & 0 & 0 & 0 & 1 & 0 & \dots \\
            & \hbox to 0pt{\dots\hss} &
\end{array}
$$
In formula (\ref{eq:stat2}) we actually computed the orthogonal 
component of the goal to the subspace spanned by $n$ share
vectors.

\section{An esoteric scheme}\label{sec:esoteric}

In this section we present a scheme which can be best described
as one which defies our intuition what a probabilistic scheme should be.

As the full set of participants is always qualified, one might be tempted to
define the secret as a function of the collective set of shares assigned to
the participants. Everyone together should be able to determine the secret,
so why the dealer bothers with determining the secret separately rather than
computing it from the shares she assigned to the participants? Scheme
\ref{scheme:non-measurable-secret} shows that this approach might lead to
problems.

\begin{scheme}\label{scheme:non-measurable-secret}
Let us split the unit interval $[0,1)$ into countably many subsets $X_i$
indexed by the set of integers $\mathbb Z$ such that all $X_i$ has outer
measure $1$ and inner measure zero. Let moreover fix the positive 
probabilities $p_j$ for $j\in\mathbb Z$ with $\sum_{j\in\mathbb Z} p_j=1$.
In this scheme there are two participants: $P=\{a,b\}$. Participant $a$ 
receives a uniform random real $r$ from the unit interval, and participant 
$b$ receives the integer $j\in\mathbb Z$ with probability $p_j$. Finally the
dealer computes the secret $s\in\mathbb Z$ as follows: she finds the index
$i\in\mathbb Z$ for which $r\in X_i$, and then sets $s=i+j$.
\end{scheme}

\noindent
From the scheme it is clear that recovering the secret participant $a$ uses
only the index $i$ for which his number is in the set $X_i$. So why don't we
give him this index rather than the uniform random real $r$? The answer is
that this index $i$ {\it has no probability distribution}. In fact, in this 
scheme the secret is {\it not a random variable}, thus the scheme is
not a probabilistic scheme at all!

\medskip
It is worth to remark that in examples \ref{scheme:ChorKush},
\ref{scheme:uniform-real}, \ref{scheme:weak-ramp}, and in others, the dealer
determined first some (or all) of the shares, and then using those values
computed the secret. What distinguishes those schemes from scheme
\ref{scheme:non-measurable-secret} is that the function determining the
secret from the shares is {\em measurable} is the former cases, and is {\em
not measurable} in the latter one.

\bibliographystyle{plain}
\bibliography{bib}

\begin{thebibliography}{10}

\bibitem{delarue}
Jacques Azéma, Marc Yor, Paul Meyer, and Thierry de~la Rue.
\newblock Espaces de {L}ebesgue.
\newblock In {\em Séminaire de Probabilités XXVII}, volume 1557 of {\em
  Lecture Notes in Mathematics}, pages 15--21. Springer Berlin / Heidelberg,
  1993.
\newblock 10.1007/BFb0087958.

\bibitem{blakley-swanson}
G.~R. Blakley and Laif Swanson.
\newblock Infinite structures in information theory.
\newblock In {\em CRYPTO}, pages 39--50, 1982.

\bibitem{ChorKush}
B.~{Chor} and E.~{Kushilevitz}.
\newblock Secret sharing over infinite domain.
\newblock {\em Journal of Cryptology}, 6(2):97--86, 1993.

\bibitem{cramer-vss}
Ronald Cramer, Ivan Damg{\aa}rd, and Stefan Dziembowski.
\newblock On the complexity of verifiable secret sharing and multiparty
  computation.
\newblock In {\em STOC}, pages 325--334, 2000.

\bibitem{infinite-secret-theory}
Laszlo Csirmaz.
\newblock Probabilistic infinite secret sharing.
\newblock 2012.
\newblock manuscript.

\bibitem{csirmaz-tardos-ligeti}
Laszlo Csirmaz, Peter Ligeti, and Gabor Tardos.
\newblock On infinite secret sharing schemes.
\newblock In {\em 10th Central Europen Ceonference on Cryptology}, Bedlewo,
  Poland, June 10--12 2010.

\bibitem{haezendonck}
J~Haezendonck.
\newblock Abstract {L}ebesgue--{R}okhlin spaces.
\newblock {\em Bulletin de la Societe Mathematique de Belgique}, 25:243--258,
  1973.

\bibitem{Ito}
M.~Itoh, A.~Saito, and T.~Nishizeki.
\newblock Secret sharing scheme realizing general access structure.
\newblock In {\em IEEE Globecom}, pages 99--102, 1987.

\bibitem{janson1997gaussian}
S.~Janson.
\newblock {\em Gaussian Hilbert Spaces}.
\newblock Cambridge Tracts in Mathematics. Cambridge University Press, 1997.

\bibitem{kallenberg}
O.~Kallenberg.
\newblock {\em Foundations of Modern Probability}.
\newblock Probability and Its Applications Series. Springer, 2010.

\bibitem{spanprogram}
Mauricio Karchmer and Avi Wigderson.
\newblock On span programs.
\newblock In {\em Structure in Complexity Theory Conference}, pages 102--111,
  1993.

\bibitem{kholevo:2001}
A.~S. Kholevo.
\newblock Sufficient statistics.
\newblock In Michiel Hazewinkel, editor, {\em Encyclopedia of Mathematics}.
  Springer Berlin / Heidelberg, 2001.

\bibitem{kunen1983set}
K.~Kunen.
\newblock {\em Set Theory: An Introduction to Independence Proofs}.
\newblock Studies in Logic and the Foundations of Mathematics. Elsevier, 1983.

\bibitem{Makar:1980}
Boshra~H. Makar.
\newblock Transfinite cryptography.
\newblock {\em Cryptologia}, 4(4):230--237, October 1980.

\bibitem{patarin:infinite}
Jacques Patarin.
\newblock Transfinite cryptography.
\newblock {\em IJUC}, 8(1):61--72, 2012.
\newblock also avaiable as \url{http://eprint.iacr.org/2010/001}.

\bibitem{Vaudenay:09}
Raphael Phan and Serge Vaudenay.
\newblock On the impossibility of strong encryption over $\aleph_0$.
\newblock In Yeow Chee, Chao Li, San Ling, Huaxiong Wang, and Chaoping Xing,
  editors, {\em Coding and Cryptology}, volume 5557 of {\em Lecture Notes in
  Computer Science}, pages 202--218. Springer Berlin / Heidelberg, 2009.

\bibitem{regev}
Oded Regev.
\newblock New lattice based cryptographic constructions.
\newblock In {\em In Proceedings of the 35th ACM Symposium on Theory of
  Computing}, pages 407--416. ACM, 2003.

\bibitem{rokhlin}
Vladimir~A. Rokhlin.
\newblock On the fundamental ideas of measure theory.
\newblock {\em Translations (American Mathematical Society)}, 10:1–54, 1962.

\bibitem{Shamir}
A.~{Shamir}.
\newblock How to share a secret.
\newblock {\em Communications of the ACM}, 22(11):612--613, 1979.

\bibitem{sperner}
Emanuel Sperner.
\newblock Ein satz über untermengen einer endlichen menge.
\newblock {\em Mathematische Zeitschrift}, 27:544--548, 1928.
\newblock 10.1007/BF01171114.

\end{thebibliography}

\end{document}